\documentclass[aps,prd,superscriptaddress,twoside,showpacs,twocolumn,nofootinbib,10pt,floatfix,showpacs]{revtex4-1}%
\usepackage{verbatim}
\usepackage{bm}
\usepackage{tikz}
\usepackage{multirow}
\usepackage{amsmath}
\usepackage{gensymb}
\usepackage{graphicx}
\usepackage{epsfig}
\usepackage{braket}
\usepackage{subfigure}
\usepackage{ulem}
\usepackage{makecell}
\usepackage{fdsymbol}

\allowdisplaybreaks

\begin{document}
\title{Ground states of all mesons and baryons in a quark model with Hidden Local Symmetry}
\author{Bing-Ran He}
\email[E-mail: ]{hebingran@njnu.edu.cn}
\affiliation{
	Department of Physics, Nanjing Normal University, Nanjing 210023, PR China
}
\author{Masayasu Harada}
\email[E-mail: ]{harada@hken.phys.nagoya-u.ac.jp}
\affiliation{
	Department of Physics, Nagoya University, Nagoya, 464-8602, Japan
}
\affiliation{
	Kobayashi-Maskawa Institute for the Origin of Particles and the Universe, Nagoya University, Nagoya, 464-8602, Japan
}
\affiliation{
	Advanced Science Research Center, Japan Atomic Energy Agency, Tokai 319-1195, Japan
}
\author{Bing-Song Zou}
\email[E-mail: ]{zoubs@itp.ac.cn}
\affiliation{
	CAS Key Laboratory of Theoretical Physics, Institute of Theoretical Physics, Chinese Academy of Sciences, Beijing 100190, China}
\affiliation{School of Physical Sciences, University of Chinese Academy of Sciences, Beijing 100049, China}
\affiliation{School of Physics, Peking University, Beijing 100871, China}

\date{\today}
\begin{abstract}
We extend the chiral quark model for $u$, $d$, $c$ and $b$ quarks with vector mesons,
which we proposed in the previous analysis, to a model with the $s$ quark.
We include the nonet pseudo-scalar and vector mesons together with the singlet scalar meson based on the SU(3)$_L \times$SU(3)$_R$ chiral symmetry combined with the Hidden Local Symmetry, which mediate force among $u$, $d$ and $s$ quarks. 
We fit the model parameters to the known ground state mesons and baryons.
We show that the mass spectra of those hadrons are beautifully reproduced.
We predict the masses of missing ground states, one meson and twenty baryons, which will be tested in the future experiment.

\end{abstract}
\maketitle

%##########
\section{\label{sec:intro}Introduction}

The ground states of mesons and baryons are the most compact particles in quantum chromodynamics (QCD), they play a crucial role in enhancing our understanding of the strong interaction among quarks. 
So far, twenty-one ground states of mesons and twenty-four ground states of baryons have been experimentally confirmed. 
The validity of a model in describing these ground states is a crucial test. 
There are, however, still one meson and twenty baryons that have not been confirmed by experiments. 
The $B_c^*$ meson is the last missing particle in the ground state of mesons. The missing baryons can be classified into three categories: (1) singly heavy baryons, with only  $\Omega_b^*$ remaining; (2) doubly heavy baryons, including $\Xi_{QQ}$, $\Xi_{QQ}^{'(*)}$, $\Omega_{QQ}$ and $\Omega_{QQ}^{'(*)}$; (3) triply heavy baryons, namely $\Omega_{QQQ}$ and $\Omega_{QQQ}^*$. Here $Q$ represent $c$ or $b$ quark. 
Numerous literatures have extensively studied these missing states, as can be found in the review articles of Refs.~\cite{Gasiorowicz:1981jz,Korner:1994nh,Kiselev:2001fw,Swanson:2006st,Vijande:2012mk,Crede:2013kia,Chen:2016spr,Wang:2020avt,Ding:2022ows} and reference therein.

Recently, Ref.~\cite{He:2023ucd} proposed a chiral quark model with inclusion of vector mesons based on the hidden local symmetry (HLS)~\cite{Meissner:1987ge,Bando:1987br,Harada:2003jx}, in addition to the scalar and pseudoscalar mesons and color contribution. 
Several hadrons including ground states constructed from $u$, $d$, $c$, $b$ quarks are studied, and it was shown that, in particular, the spectra of baryons including good diquark are dramatically improved by the inclusion of the vector meson contribution.

In this paper we extend the model to include the strange quark and study the mass spectra of hadrons including the strange quark. 
In the present chiral quark model phenomenology, various mesons play distinct roles: (1) the exchange of $\pi$ mesons leads to the $\pi-\rho$ splitting~\cite{Blanco:1999qt}; (2) the exchange of $\omega$ meson resolves the ``good-diquark" problem that arises when constructing baryons from two light quarks~\cite{He:2023ucd}; (3) 
the exchange of $K$ mesons contributes to the $\eta-\eta'$~\cite{Vijande:2004he}, $\Lambda-\Sigma$, $\Xi_c-\Xi_c'$ and $\Xi_b-\Xi_b'$ splittings.   
To ensure a comprehensive framework that incorporates all these meson exchanges consistently, we incorporate the effects of meson exchange using a nonet of pseudo-scalar and vector mesons based on the HLS formalism. 
We will show that the masses of existing ground states are beautifully reproduced by a suitable choice of the model parameters.
Then, we predict mass spectra of missing ground states.
We expect that, among the missing ground states, 
the prediction on the $\Omega_b^*$ will be a crucial test of the model, 
since the quark composition of $\Omega_b^*$ is $ssb$, so the force caused by mesonic exchange will affect its mass. 
On the other hand, the majority of other missing ground states are primarily governed by the color force resulting from one-gluon exchange (OGE) and the confinement potential (CON). 
However, due to the absence of double beauty and beauty-charm baryons, accurately determining the coupling strengths of $b$-$b$ and $b$-$c$ interactions has become a critical challenge. 
Achieving a systematic description of the missing ground states requires striking a balance between the mesonic and color potentials. 

This paper is organized as follows:
In Sec.~\ref{sec:model}, we introduce a new chiral model which includes nonet pseudo-scalar and vector mesons together with a flavor singlet scalar meson.
Then, in Sec.~\ref{sec:results}, we show the numerical results.
Finally, a brief summary is given in Sec.~\ref{sec:summary}.

%##########
\section{\label{sec:model} Quark model with SU(3) Hidden Local Symmetry}
%{\bf``The model''}-

In the previous analysis, we have included pions, an iso-singlet scalar meson which expresses the two-pion contribution, iso-singlet omega meson and iso-vector rho meson, which couple to up and down quarks.
In the present analysis, we extend the model to include the strange quark. Associated with this extension, we include the following mesons into the model.
(1)~Pseudoscalar mesons:
we include all the members of nonet pseudoscalar mesons $\pi$, $K$, $\eta$ and $\eta'$ as an extension of the previous analysis. 
We note that the $\eta$ and $\eta'$ mesons provide contributions to the up and down quarks which are not included in the previous analysis, so that we will refit hadrons with strange quarks.~
(2)~Vector mesons: we include nonet vector mesons $\omega$, $\rho$, $K^\ast$ and $\phi$.~
(3)~Scalar mesons: 
we include a flavor singlet scalar meson as an extension of the iso-singlet scalar meson included in the previous analysis.

Now, the Hamiltonian is written as: 
\begin{eqnarray}\label{eq:H}
	H&=&\sum_{i=1}\left(m_i+\frac{p_i^2}{2m_i}\right)-T_{CM} + \sum_{j>i=1}\left(V^{\rm CON}_{ij}+V^{\rm OGE}_{ij}\right.\nonumber\\
	&&\left.+V_{ij}^{\bar{\sigma}}+V_{ij}^{PS}
	+V_{ij}^{V}\right)\, ,
\end{eqnarray}
where $m_{i}$ and $p_{i}$ are the mass and the momentum of $i$-{th} quark, $T_{CM}$ is the kinetic energy of the center of mass of the system. 
$V^{\rm CON}_{ij}$ and $V^{\rm OGE}_{ij}$ represent the gluonic potential of confinement and one-gluon-exchange, which are given in~\cite{Vijande:2004he}.  $V_{ij}^{\bar{\sigma}}$, $V_{ij}^{PS}$ and $V_{ij}^{V}$ represent scalar, pseudo-scalar and vector potential, respectively.
The pseudo-scalar and vector potentials are decomposed as
\begin{eqnarray}
&&V_{ij}^{PS}=V_{ij}^{\eta}+V_{ij}^{\eta'}
+V_{ij}^{\pi}+V_{ij}^{K}\,,\nonumber\\
&&V_{ij}^{V}=V_{ij}^{\omega}+V_{ij}^{\phi}
+V_{ij}^{\rho}+V_{ij}^{K^*}\, , 
\end{eqnarray}
where
$V_{ij}^{\eta}$, $V_{ij}^{\eta'}$, $V_{ij}^{\pi}$ and $V_{ij}^{K}$ represent the potentials generated by the exchanges of $\eta$, $\eta'$,  $\pi$ and $K$ mesons, respectively, while  
$V_{ij}^{\omega}$, $V_{ij}^{\phi}$, $V_{ij}^{\rho}$ and $V_{ij}^{K^*}$ are by $\omega$, $\phi$, $\rho$ and $K^*$ mesons. 
Their explicit forms are given as 
\begin{eqnarray}  &&V_{ij}^{\bar\sigma}=V_{ij}^{s=\bar\sigma,g_s=g_{\bar\sigma q}}\lambda_i^q\lambda_j^q
	+ V_{ij}^{s=\bar\sigma,g_s=g_{\bar\sigma s}}\lambda_i^s\lambda_j^s\,,\nonumber\\ 
	&&V_{ij}^{\eta}=V_{ij}^{p=\eta,g_p=g_{\eta q}}\lambda_i^{q}\lambda_j^{q}
	+
	V_{ij}^{p=\eta,g_p=g_{\eta s}}\lambda_i^{s}\lambda_j^{s}\,,\nonumber\\
	&&V_{ij}^{\eta'}=V_{ij}^{p=\eta',g_p=g_{\eta'q}}\lambda_i^{q}\lambda_j^{q
	}+
	V_{ij}^{p=\eta',g_p=g_{\eta' s}}\lambda_i^{s}\lambda_j^{s}\,,\nonumber\\
	&&V_{ij}^{\pi}=V_{ij}^{p=\pi,g_p=g_{\pi}}\sum_{a=1}^3\lambda_i^a\lambda_j^a\,,\nonumber\\
	&&V_{ij}^{K}=V_{ij}^{p=K,g_p=g_K}\sum_{a=4}^7\lambda_i^a\lambda_j^a\,,\nonumber\\
	&&V_{ij}^{\omega}=V_{ij}^{v=\omega,g_v=g_{\omega q}}\lambda_i^q\lambda_j^q
	+V_{ij}^{v=\omega,g_v=g_{\omega s}}\lambda_i^s\lambda_j^s\,,\nonumber\\
	&&V_{ij}^{\phi}=
	V_{ij}^{v=\phi,g_v=g_{\phi q}}\lambda_i^q\lambda_j^q+
	V_{ij}^{v=\phi,g_v=g_{\phi s}}\lambda_i^s\lambda_j^s\,,\nonumber\\
	&&V_{ij}^{\rho}=V_{ij}^{v=\rho,g_v=g_{\rho}}\sum_{a=1}^3\lambda_i^a\lambda_j^a\,,\nonumber\\
	&&V_{ij}^{K^*}=V_{ij}^{v=K^*,g_v=g_{K^* }}\sum_{a=4}^7\lambda_i^a\lambda_j^a\,.
\end{eqnarray} 
Here $\lambda^a$ ($a=1,2,\ldots,7$) are flavor $SU(3)$ Gell-Mann matrices, $\lambda^q$ and $\lambda^s$ are expressed as
\begin{eqnarray}
	\lambda^q=
	\begin{pmatrix}
		1&0&0\\
		0&1&0\\
		0&0&0
	\end{pmatrix}
	\,,\,
	\lambda^s=
	\begin{pmatrix}
		0&0&0\\
		0&0&0\\
		0&0&1
	\end{pmatrix}
	\,.
\end{eqnarray} 
$V^s$, $V^{p}$ and $V^v$ are common parts of scalar, pseudo-scalar and vector mesons, respectively, which are given in~\cite{He:2023ucd}. 

The mesonic potentials for $q\bar{q}$ ($q=u,d$) is obtained by performing a G-parity transformation of that in $q{q}$ case. In the case that G parity is not well defined, e.g., $K$ and $K^*$, the transform is given by $\lambda_i^a\lambda_j^a\to \lambda_i^a(\lambda_j^a)^*$.

We treat the mass of ${\bar\sigma}$ 
as a model parameter, while the mass of $\eta$, $\eta'$, $\pi$, $K$, $\omega$, $\phi$, $\rho$ and $K^*$ are used as their PDG values. 
For the coupling constants of those mesons to quarks, we require the following relation based on the SU(3) flavor symmetry.
\begin{enumerate}
\item 
pseudo-scalar mesons:  
$g_{\eta s}=g_{\eta q}-\sqrt{3}\cos\theta_p g_{\pi}$,
$g_{\eta' q}=- \cot \theta_p g_{\eta q}+\frac{1}{\sqrt{3}\sin \theta_p}g_\pi$,  
$g_{\eta' s}=- \cot \theta_p g_{\eta q}+\frac{\cos \theta_p\cot \theta_p-2\sin \theta_p}{\sqrt{3}}g_\pi$ and $g_\pi=g_K$. Here $\theta_p=-11.3^\circ$ is taken from PDG. 

\item 
vector meson: 
$g_{\omega s}=g_{\omega q}-g_\rho$, 
$g_{\phi q}= -\sqrt{\frac{1}{2}}(g_{\omega q}- g_\rho)$,
$g_{\phi s}=-\sqrt{\frac{1}{2}}(g_{\omega q}+ g_\rho)$, $g_\rho=g_{K^*}$ and $f_{\omega s}=f_{\omega q}-f_\rho$, 
$f_{\phi q}= -\sqrt{\frac{1}{2}}(f_{\omega q}- f_\rho)$,
$f_{\phi s}=-\sqrt{\frac{1}{2}}(f_{\omega q}+ f_\rho)$, $f_\rho=f_{K^*}$.

\end{enumerate}

%%##########
\section{\label{sec:results}
Numerical Results}

We solve two and three body problems for  mesons and baryons 
by using Gaussian expansion method(GEM)~\cite{Hiyama:2003cu}. 
We determine the model parameters by minimizing the $\chi^2$ of the system defined by 
\begin{eqnarray}
	\chi^2=\sum_{i}\left(\frac{m_i({\rm the})-m_i({\rm exp})}{{\rm Err_i(sys)}}\right)^2\,,
\end{eqnarray}
where $m_i({\rm the})$ and $m_i({\rm exp})$ are theoretical and experimental mass of each particle, respectively. 
The system error is determined as  
\begin{eqnarray}
{\rm Err(sys)}=\sqrt{{\rm Err(exp)}^2+{\rm Err(the)}^2}\,,
\end{eqnarray}
where
${\rm Err(exp)}$ is the experimental error taken from PDG~\cite{ParticleDataGroup:2022pth}, while ${\rm Err(the)}$ represents
the model limitation error as we do not include isospin breaking effects, and also ignore mixing effects like S-D and P-F mixings between meson states.
We take ${\rm Err(the)}$ as $40$\,MeV for ground-state and $80$\,MeV for excited-state of mesons and baryons. 
In the present study we use 21 meson ground states and one meson excited state $\Upsilon(2s)$, together with 24 baryon ground states, totally 46 hadron states as inputs. 
We take $\eta$-$\eta'$ mixing parameter $\theta_p$ and 8 masses of pseudoscalar and vector mesons which contribute to the potential from PDG so the number of free parameter (including cutoffs) is $22=31-9$. 
As a result, the degree of freedom (dof) is $24=46-22$. 

In Table~\ref{tab:para}, we list the values of model parameters for best fitted case,
where $\chi^2/{\rm dof}=8.8/24$ leading to the reliability of present analysis as about $99.8\%$.
\begin{table}[htp]
\caption{Best fitted values of model parameters, where $\chi^2/{\rm dof}=8.8/24$ leading to the reliability of present analyze as about $99.8\%$. 
	}\label{tab:para}
	\begin{ruledtabular}
		\begin{tabular}{ c c  c c}
			$m_u = m_{d}({\rm MeV})$              & 381.1   & $m_{\bar\sigma}({\rm fm}^{-1})$                                                                  & 2.53   \\
			$m_s({\rm MeV})$                      & 551.6   & $m_{\eta}({\rm fm}^{-1})$                                                                        & 2.78   \\
			$m_c({\rm MeV})$                      & 1735.0  & $m_{\eta'}({\rm fm}^{-1})$                                                                       & 4.85   \\
			$m_b({\rm MeV})$                      & 5094.8  & $m_{\pi}({\rm fm}^{-1})$                                                                         & 0.7    \\
			$a_c({\rm MeV})$                      & 352.7   & $m_{K}({\rm fm}^{-1})$                                                                           & 2.51   \\
			$\mu_c({\rm fm}^{-1})$                & 3.3     & $m_{\omega}({\rm fm}^{-1})$                                                                      & 3.97   \\
			$\Delta({\rm MeV})$                   & 327.6   & $m_{\phi}({\rm fm}^{-1})$                                                                        & 5.17   \\
			$\alpha_0$                            & 0.703   & $m_{\rho}({\rm fm}^{-1})$                                                                        & 3.93   \\
			$\Lambda_0({\rm fm}^{-1})$            & 0.835   & $m_{K^*}({\rm fm}^{-1})$                                                                         & 4.54   \\
			$\mu_0({\rm MeV})$                    & 300.688 & $\theta_p(^\circ)$                                                                               & -11.3  \\
			$\hat{r}_0({\rm MeV} \cdot {\rm fm})$ & 23.067  & \makecell{$\Lambda_{\bar\sigma} = \Lambda_{\pi}= \Lambda_{\eta}$\\$=\Lambda_{K}({\rm fm}^{-1})$} & 4.2    \\
			$g_{\bar\sigma q}$=$g_{\bar\sigma s}$ & -0.003  & $\Lambda_{\omega}=\Lambda_{\rho}=\Lambda_{K^*}({\rm fm}^{-1})$                                   & 5.2    \\
			$g_{\pi}=g_{K}$                       & 2.912   & $ \Lambda_{\eta'}=\Lambda_{\phi}({\rm fm}^{-1})$                                                 & 6.2    \\
			$g_{\eta q}$                          & 1.177   &                                                                                                  $f_{\omega q}$                                                                                   & 0.42        \\
			$g_{\omega q}$                        & 1.118   & $f_{\rho}=f_{K^*}$                                                                               & -0.222   \\
			$g_{\rho}=g_{K^*}$                    & -0.323  &                                                                               & 
		\end{tabular}
	\end{ruledtabular}
\end{table}
Here, to understand Table~\ref{tab:para} we clarify the operators included in the potentials with the sign of each contribution, which are summarized in Table~\ref{tab:oper}.  
\begin{table}[htp]
	\caption{\label{tab:oper}List of operators included in the potentials for $L=0$ states and the sign of the contributions.
		The left (right) side of slash represents $qq$ ($q\bar{q}$), respectively.
	}
	\begin{ruledtabular}
		\begin{tabular}{ c c  c c c}
			& 1     & $\lambda_i\lambda_j$ & $\sigma_i\sigma_j$ & $\sigma_i\sigma_j\cdot\lambda_i\lambda_j$ \\
			$\bar\sigma$ & $-/-$ &                &                    &                                     \\ 
			$\eta/\eta'$    &       &                & $+/+$                   &                                \\
			$\pi/K$    &       &                &                    & $+/-$                               \\
			$\omega/\phi$           & $+/-$ &                & $+/-$              &                                     \\
			$\rho/K^*$            &       & $+/+$          &                    & $+/+$ \\
			OGE      & $-/-$ &                & $+/+$              &                                     \\
			CON      & $+/+$ &                &                    &                                     
		\end{tabular}
	\end{ruledtabular}
\end{table}
Combining Tables~\ref{tab:para} and \ref{tab:oper}, we observe that the most significant meson exchanges are $\pi/K$ and $\omega/\phi$. 
This observation is understandable since these mesons play opposite roles in the $qq$ and $q\bar{q}$ sectors. 
$\rho/K^*$ mesons make minor modifications to the $\pi/K$ mesons in the $\sigma_i \sigma_j \cdot \lambda_i \lambda_j$ sector, and they have little impact on the $\lambda_i \lambda_j$ sector. 
Forces mediated by $\bar{\sigma}$ and $\eta/\eta'$ mesons exhibit similar characteristics to the $V^{\rm OGE}$ and $V^{\rm CON}$, albeit with modified strength or effective range.

\begin{figure}[htp]
	\centering
\includegraphics[scale=0.21]{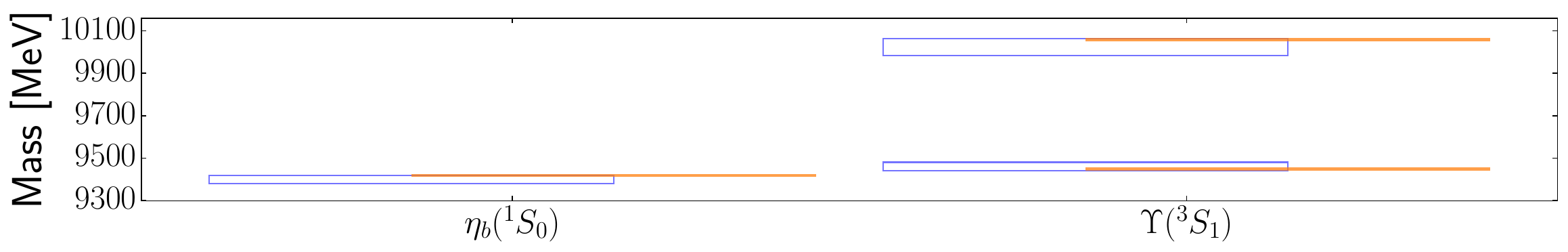}
\includegraphics[scale=0.21]{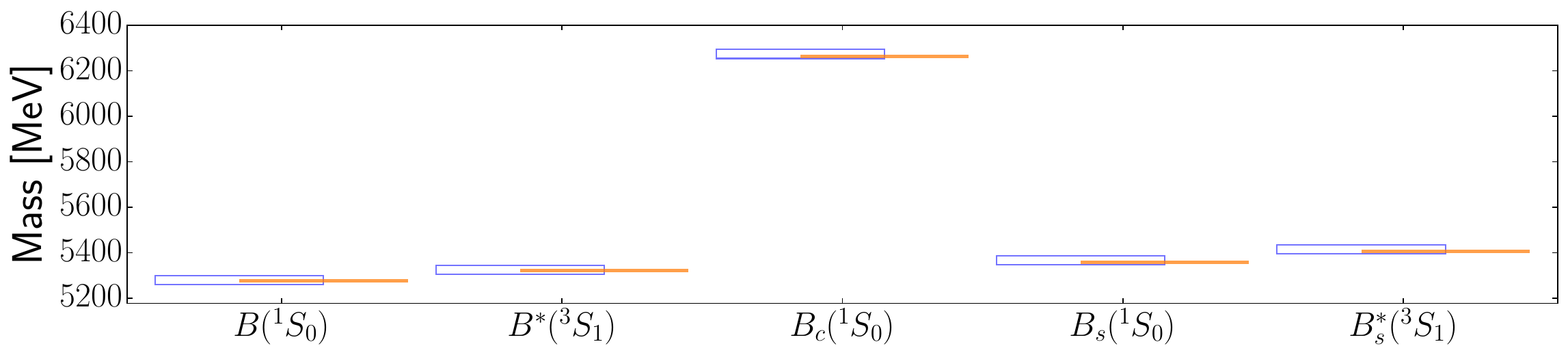}
\includegraphics[scale=0.21]{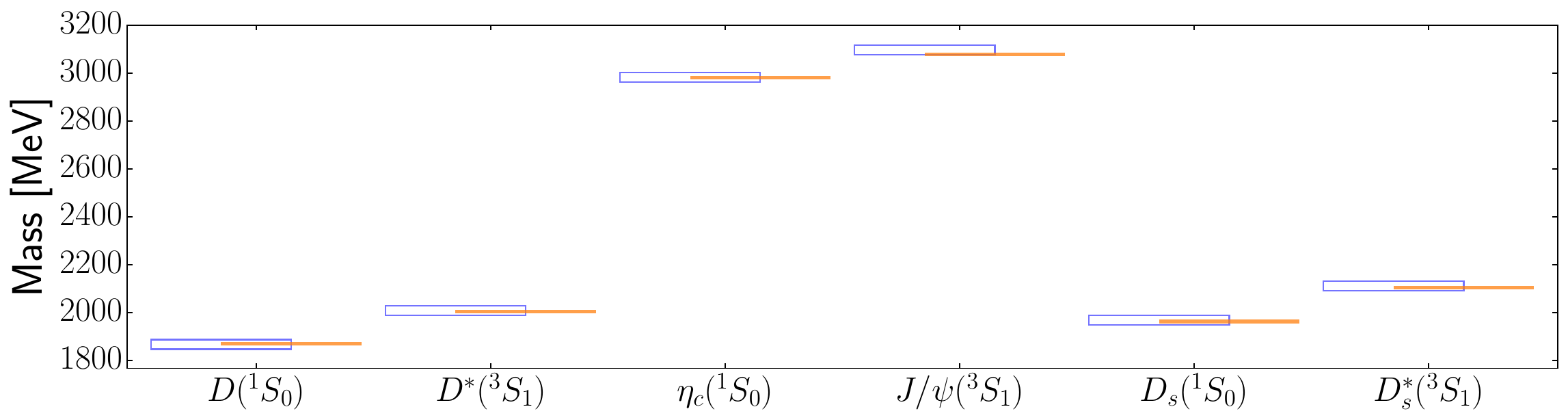}
\includegraphics[scale=0.21]{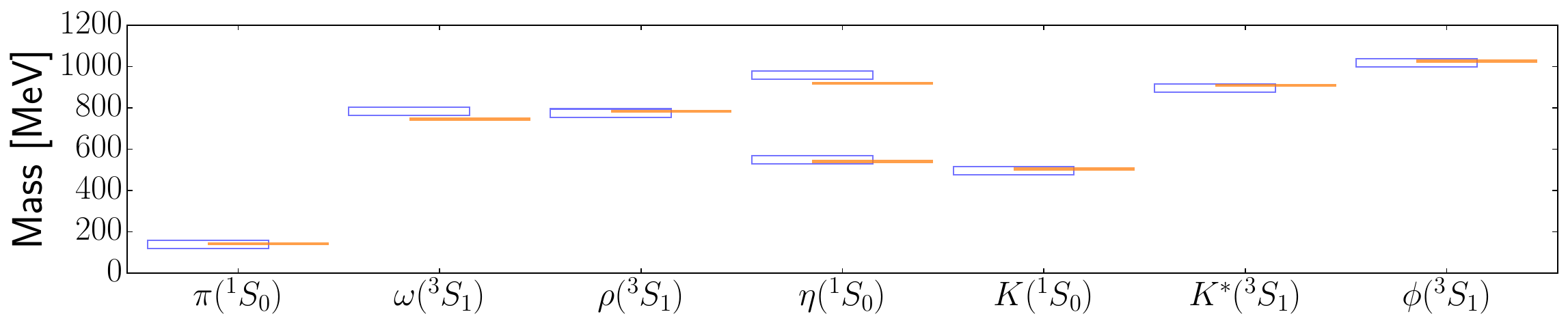}
	\caption{Mass spectrum of mesons. Blue boxes represent $m({\rm exp})\pm{\rm Err(sys)}$, while the orange lines represent the predicted masses.}
	\label{fig:meson_mass_1}
\end{figure}

\begin{figure}[htp]
	\centering
\includegraphics[scale=0.21]{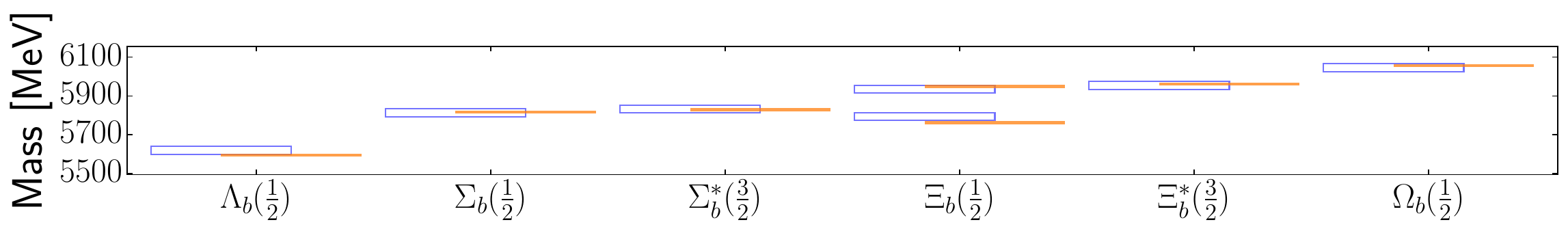}
\includegraphics[scale=0.21]{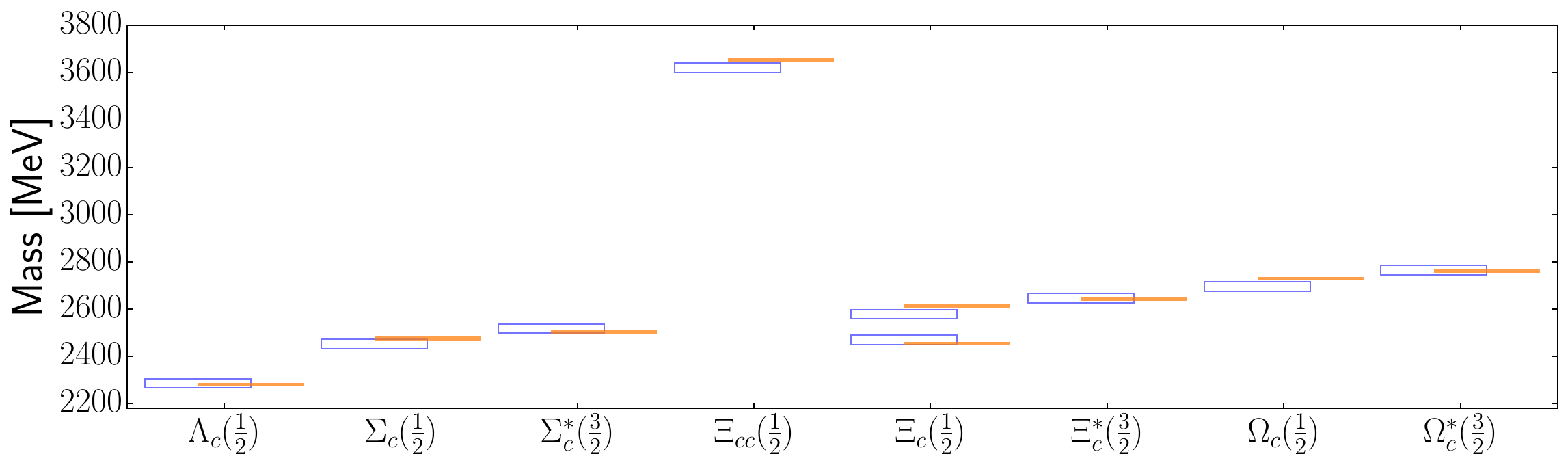}
\includegraphics[scale=0.21]{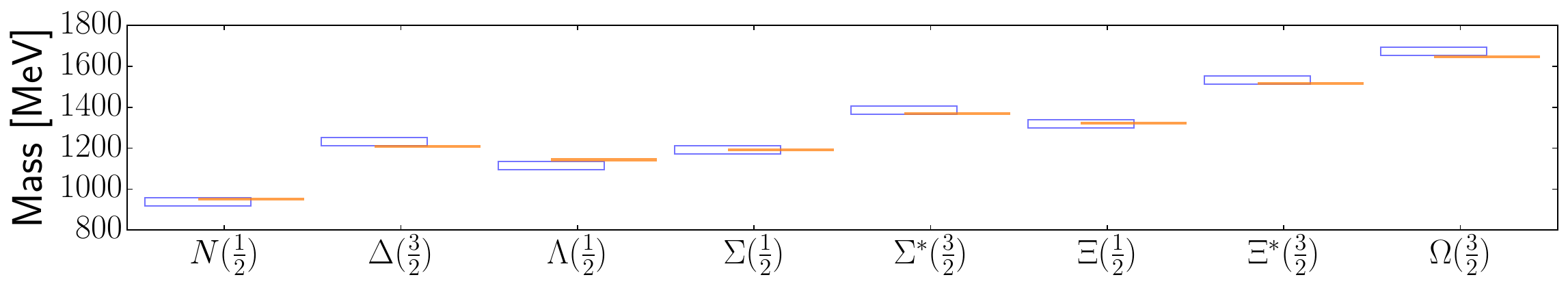}
	\caption{Mass spectrum of baryons. The colors have the same meaning as shown in Fig.~\ref{fig:meson_mass_1}.}
	\label{fig:baryon_mass_1}
\end{figure}

In Figs.~\ref{fig:meson_mass_1} and \ref{fig:baryon_mass_1}, we show mass spectrum of mesons and baryons, respectively, obtained by using best fitted model parameters.
These figures show that all the existing ground states of mesons and baryons are beautifully reproduced.

\begin{table}[htp]
	\caption{\label{tab:mass_spec} Predicted mass spectrum (in MeV) of ground states for mesons and baryons.
	}
	\begin{ruledtabular}
		\begin{tabular}{ c c  c c c c c c }
$\pi$ & $\omega$ & $\rho$ & $\eta$ & $\eta'$ & $K$ & $K^*$ & $\phi$ \\
141.3 & 745.5 &783.4 & 541.1& 918.7& 503.3 & 909.8
& 1027.6\\
\hline
$D$ & $D^*$ & $\eta_c$ & $J/\Psi$& $D_s$ & $D_s^*$ & &  \\
1869.4 &2005 &2981.9 & 3078& 1963& 2103.8& & \\
\hline
$B$ & $B^*$ & $\eta_b$ & $\Upsilon$& $B_s$ & $B_s^*$ & $B_c$ & $\Upsilon(2S)$ \\
5277.6 &5321.9 & 9417.8& 9448.2& 5358.9 & 5406.2&6263.5 & 10059.1\\
\hline
$N$ & $\Delta$ & $\Lambda$ & $\Sigma$ & $\Sigma^*$ & $\Xi$ & $\Xi^*$ & $\Omega$ \\
951.1 & 1208.9& 1143& 1192.5& 1368.6& 1322.4& 1516.5& 1647.5\\
\hline
$\Lambda_c$ & $\Sigma_c$ & $\Sigma_c^*$ & $\Xi_{c}$ & $\Xi_{c}'$ & $\Xi_{c}^*$ & $\Omega_c$ & $\Omega_c^*$ \\
2279.6 & 2476& 2504.7& 2455.2& 2614.5& 2642.5& 2729& 2760.4\\
\hline
$\Lambda_b$ & $\Sigma_b$ & $\Sigma_b^*$ & $\Xi_{b}$ & $\Xi_{b}'$ & $\Xi_{b}^*$ & $\Omega_b$ & $\Xi_{cc}$ \\
5594.9 & 5818& 5829.7& 5762.3& 5949.1& 5960.6& 6055.1& 3653.9\\
\hline
$B_c^*$&$\Xi_{cc}^*$ &$\Xi_{bc}$ & $\Xi_{bc}'$ & $\Xi_{bc}^*$ & $\Xi_{bb}$ & $\Xi_{bb}^*$ &   \\
6306.9 &3698.9 & 6943.9&6958.4 &6976.4 & 10176& 10198.5& \\
\hline
$\Omega_{cc}$&$\Omega_{cc}^*$ &$\Omega_{bc}$ & $\Omega_{bc}'$ & $\Omega_{bc}^*$ & $\Omega_{bb}$ & $\Omega_{bb}^*$ & \\
3763.9 & 3806.8&7044.7 &7057.7 & 7076&10266.9 & 10289& \\
\hline
$\Omega_{b}^*$ & $\Omega_{ccc}$ & $\Omega_{ccb}$ & $\Omega_{ccb}^*$& $\Omega_{bbc}$& $\Omega_{bbc}^*$& $\Omega_{bbb}$& \\
6068.1 & 4795& 8014.9& 8027.5&11191.9 & 11207.7& 14351.2& \\
		\end{tabular}
	\end{ruledtabular}
\end{table}

\begin{figure}[htp]
	\centering
\includegraphics[scale=0.21]{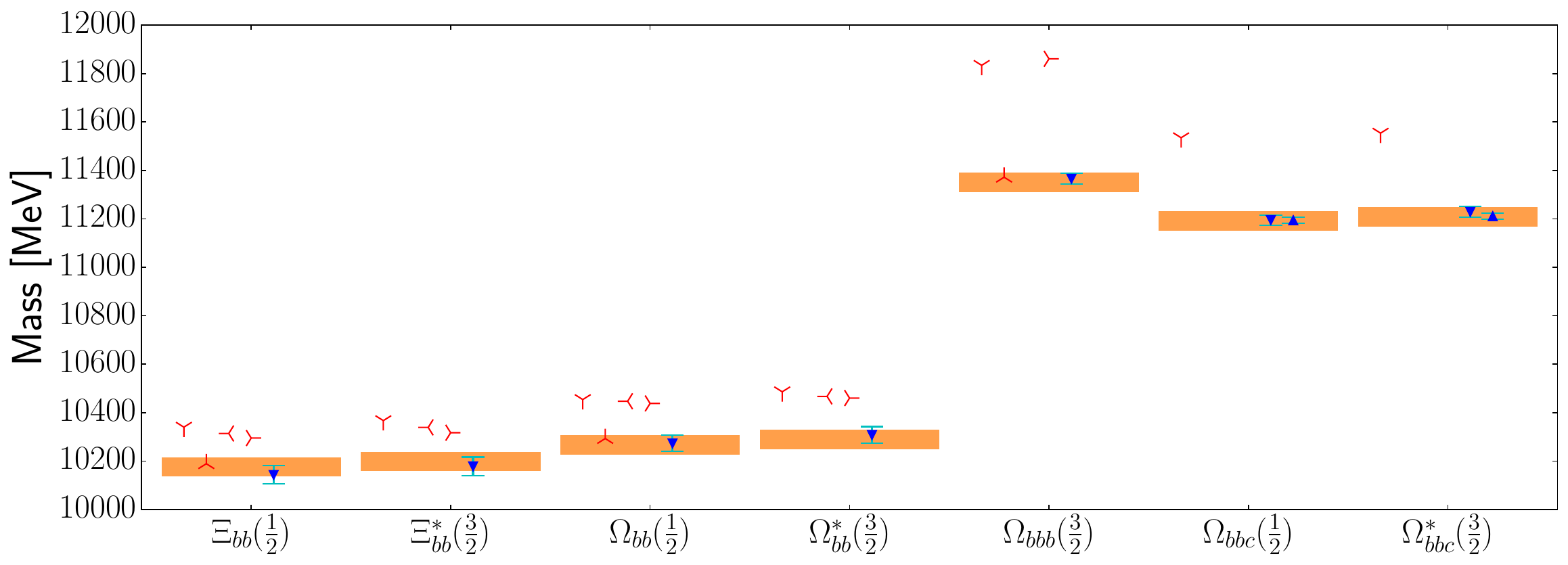}
\includegraphics[scale=0.21]{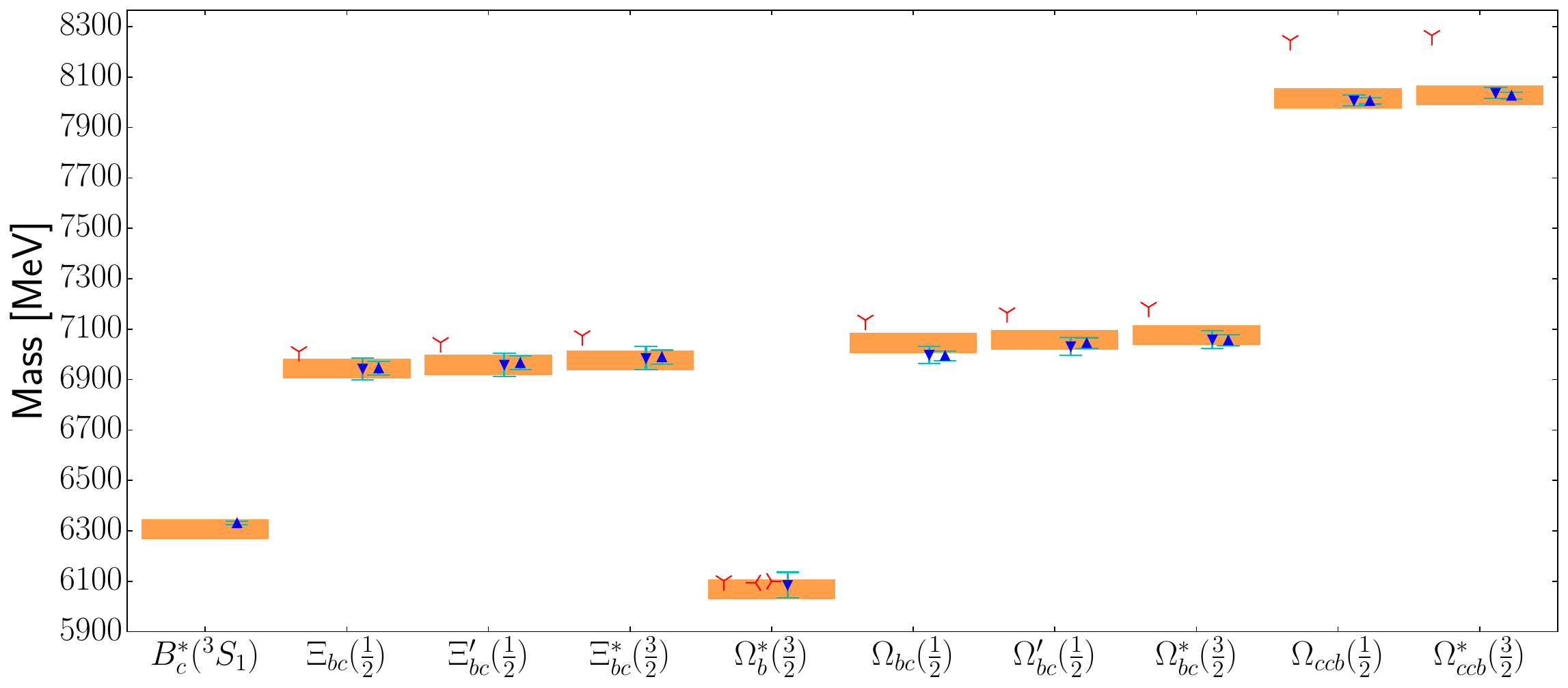}
\includegraphics[scale=0.21]{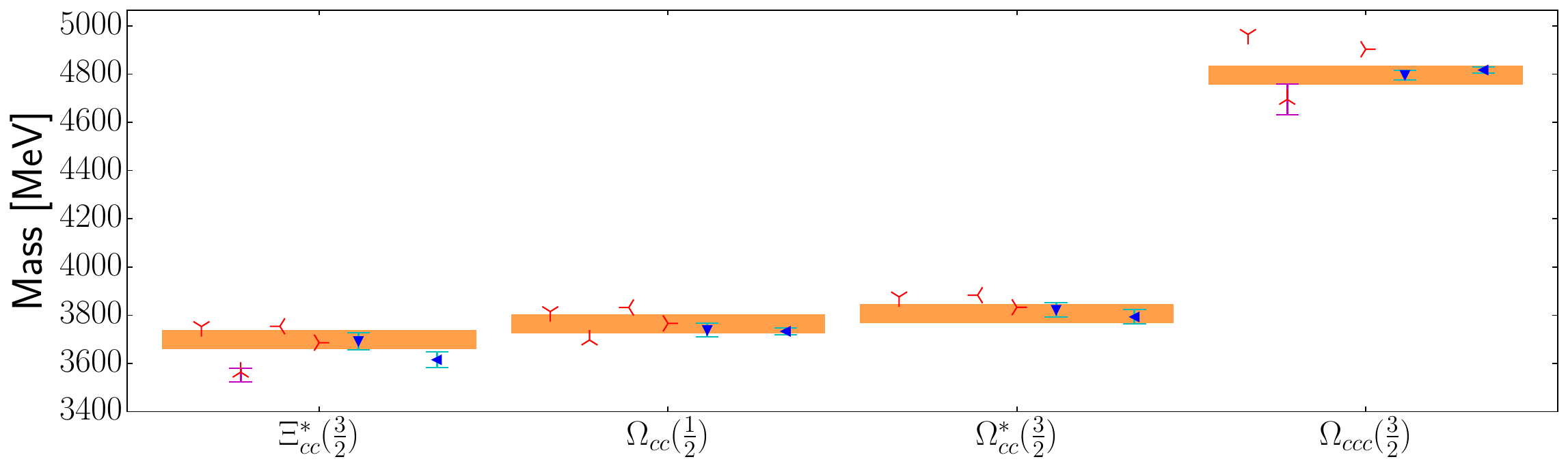}
\caption{
Predicted mass spectrum of missing meson and baryons which have not been experimentally confirmed shown by orange line with $40$\,MeV error.
Predictions of some other quark models ({\color{red}$\downY$}~\cite{Roberts:2007ni},  {\color{red}$\upY$}~\cite{Vijande:2004at,Valcarce:2008dr,Vijande:2014uma}, {\color{red}$\leftY$}~\cite{Yoshida:2015tia}, {\color{red}$\rightY$}~\cite{Ortiz-Pacheco:2023kjn}) and lattice QCD calculations ({\color{blue}$\blacktriangledown$}~\cite{Brown:2014ena}, {\color{blue}$\blacktriangle$}~\cite{Mathur:2018epb}, {\color{blue}$\blacktriangleleft$}~\cite{Bahtiyar:2020uuj}) are shown for comparison. 
The values of $\Omega_{bbb}$ shown here are shifted by $-3000$\,MeV. }
	\label{fig:missing_mass_1}
\end{figure}

We present the predicted mass spectra of mesons and baryons, including missing ground states that have not been observed experimentally, in Table.~\ref{tab:mass_spec}. 
In Fig.~\ref{fig:missing_mass_1}, we
compare our results (orange line with $40$\,MeV error) with those obtained from other quark models ({\color{red}$\downY$}~\cite{Roberts:2007ni},  {\color{red}$\upY$}~\cite{Vijande:2004at,Valcarce:2008dr,Vijande:2014uma}, {\color{red}$\leftY$}~\cite{Yoshida:2015tia}, {\color{red}$\rightY$}~\cite{Ortiz-Pacheco:2023kjn}) and lattice QCD calculations ({\color{blue}$\blacktriangledown$}~\cite{Brown:2014ena}, {\color{blue}$\blacktriangle$}~\cite{Mathur:2018epb}, {\color{blue}$\blacktriangleleft$}~\cite{Bahtiyar:2020uuj}).
It is evident that our mass spectra are consistent with lattice calculations in Refs.~\cite{Brown:2014ena,Mathur:2018epb,Bahtiyar:2020uuj}. 
In the limited comparison shown in Fig.~\ref{fig:missing_mass_1}, the state $\Omega_b^*$ is the most accurately predicted among all the models. 
This can be understood as $\Omega_b^*$ being an extension of $\Omega_b$, similar to the relation between $\Omega_c^*$ and $\Omega_c$.
When dealing with double beauty and beauty-charm quarks, many quark models encounter challenges in predicting their properties. 
This difficulty arises from the limited availability of experimental data on double beauty and beauty-charm baryons, which hampers the determination of the coupling strengths between $b$-$b$ and $b$-$c$ quarks. 
However, we can still gather some insights into the $b$-$b$ and $b$-$c$ systems by assuming that meson exchange exclusively occurs among the light $u$, $d$ and $s$ quarks. In this regard, we can draw knowledge from the $b$-$\bar{b}$ and $b$-$\bar{c}$ systems, such as the $\Upsilon$, $\eta_b$ and $B_c$ families. 
In the present analysis,
by utilizing the chiral quark model with the HLS, we can achieve a better understanding of the ground states in both experimental and lattice QCD studies.

We further predict the masses of the tetraquarks $T_{cc}$ and $T_{bb}$ are $17.9$\,MeV and $171.0$\,MeV below their corresponding thresholds of $DD^*$ and $BB^*$, respectively.  
Taking into account the systematic error of the ground state, which is approximately $40$\,MeV,  
we conclude that the biding energy for $T_{cc}$ is consistent with the experimental results, as previously predicted in Ref.~\cite{He:2023ucd}. 
We consider that the predictions of missing states and $T_{bb}$ will be tested in future experiments.

%%##########
\section{\label{sec:summary}Summary}
%{\bf ``Summary''}-

We constructed a chiral quark model, in which the nonet pseudo-scalar and vector mesons together with the singlet scalar meson are included based on the SU(3)$_L \times$SU(3)$_R$ chiral symmetry to mediate force among $u$, $d$ and $s$ quarks.  
We performed a fitting of the model parameters to the known masses of 46 hadron states, yielding a $\chi^2/{\rm dof}$ value of $0.36$. The obtained results demonstrate a remarkable agreement, with the masses of all 46 experimentally confirmed hadrons being accurately reproduced. 
Furthermore, the predictions for 21 ground states align well with the results obtained from lattice QCD analyses. 
Based on our current understanding, this paper represents a pioneering achievement in describing all 45+21 ground states of mesons and baryons using a single parameter set. Notably, we have accomplished this feat with an overall error of approximately $40$\,MeV, marking a significant milestone as it is the first instance where such comprehensive results have been successfully attained.  
The predictions concerning the masses of missing ground states, as well as tetraquarks $T_{bb}$, will be tested in future experiments.

\begin{acknowledgments}
	B.R.~He was supported in part by the National Natural Science Foundation of China (Grant Nos. 11705094 and 12047503), Natural Science Foundation of Jiangsu Province, China (Grant No. BK20171027), Natural Science Foundation of the Higher Education Institutions of Jiangsu Province, China (Grant Nos. 17KJB140011 and 22KJB140012).
	
	B.S. Zou was supported by the National Natural Science Foundation
	of China (NSFC) and the Deutsche Forschungsgemeinschaft
	(DFG, German Research Foundation) through the funds provided to
	the Sino-German Collaborative Research Center TRR110 Symmetries
	and the Emergence of Structure in QCD (NSFC Grant No.
	12070131001, and DFG Project-ID 196253076-TRR 110), the NSFC
	(11835015, and 12047503), and the Grant of Chinese Academy of
	Sciences (XDB34030000).
	
	M.~Harada was supported in part by JSPS KAKENHI Grant No. 20K03927 and 23H05439.
	
\end{acknowledgments}

%\newpage
%\clearpage

\end{document}